\documentclass[12pt]{article}

\pdfoutput=1

\usepackage{color}

\usepackage{amsmath,amssymb,latexsym}
\usepackage[utf8]{inputenc}  

\usepackage{cite}

\usepackage[T1]{fontenc}
\usepackage{times}

\usepackage{epstopdf}

\usepackage{graphics,epsfig}

\usepackage{caption}
\usepackage{fdsymbol}

\usepackage{indentfirst}

\usepackage{colortbl}

\usepackage{xcolor}

\usepackage{enumitem}

\definecolor{olive}{rgb}{0.3, 0.4, .1}
\definecolor{fore}{RGB}{249,242,215}
\definecolor{back}{RGB}{51,51,51}
\definecolor{title}{RGB}{255,0,90}
\definecolor{dgreen}{rgb}{0.,0.6,0.}
\definecolor{gold}{rgb}{1.,0.84,0.}
\definecolor{JungleGreen}{cmyk}{0.99,0,0.52,0}
\definecolor{BlueGreen}{cmyk}{0.85,0,0.33,0}
\definecolor{RawSienna}{cmyk}{0,0.72,1,0.45}
\definecolor{Magenta}{cmyk}{0,1,0,0}
\definecolor{darkred}{rgb}{0.55,0.,0.}
\definecolor{coxgreen}{rgb}{0.4,0.5,0.08}
\definecolor{dolgreen}{rgb}{0.33,0.42,0.18}

\definecolor{greym}{RGB}{76,76,76}

\definecolor{UniBlue}{RGB}{83,91,170}

\let\a=\alpha \let\b=\beta \let\g=\gamma \let\d=\delta \let\e=\epsilon
 \let\k=\kappa
\let\l=\lambda \let\m=\mu \let\n=\nu \let\x=\xi \let\p=\pi 
\let\s=\sigma 

\let\w=\omega       \let\D=\Delta \let\Th=\Theta 
\let\X=\Xi  \let\S=\Sigma  \let\Y=\Psi
 \let\W=\Omega
\let\la=\label  
  
 \def\bd{\begin{document}} \def\ed{\end{document}}
\def\ds{\documentstyle} \let\fr=\frac \let\bl=\bigl \let\br=\bigr
\let\Br=\Bigr \let\Bl=\Bigl
\let\bm=\bibitem
\let\na=\nabla
\def\tU{{\widetilde U}}
\let\pa=\partial \let\ov=\overline
\def\ie{{\it i.e.\ }}
\newcommand{\be}{\begin{equation}}
\newcommand{\ee}{\end{equation}}
\def\ba{\begin{array}}
\def\ea{\end{array}}

\def\bei{\begin{itemize}}
\def\eei{\end{itemize}}

\def\ben{\begin{enumerate}}
\def\een{\end{enumerate}}

\def\ft#1#2{{\textstyle{{\scriptstyle #1}\over {\scriptstyle #2}}}}
\def\fft#1#2{{#1 \over #2}}
\def\F#1#2{{ F_{#1}^{(#2)} }}
\def\cF#1#2{{ {\cal F}_{#1}^{(#2)} }}

\def\R{{\bf R}}
\def\sst#1{{\scriptscriptstyle #1}}
\def\oneone{\rlap 1\mkern4mu{\rm l}}
\def\e7{E_{7(+7)}}
\def\td{\tilde}
\def\wtd{\widetilde}
\def\im{{\rm i}}
\def\bog{Bogomol'nyi\ }
\newcommand{\ho}[1]{$\, ^{#1}$}
\newcommand{\hoch}[1]{$\, ^{#1}$}
\newcommand{\bea}{\begin{eqnarray}}
\newcommand{\eea}{\end{eqnarray}}
\newcommand{\ra}{\rightarrow}
\newcommand{\lra}{\longrightarrow}
\newcommand{\Lra}{\Leftrightarrow}
\newcommand{\ap}{\alpha^\prime}
\newcommand{\bp}{\tilde \beta^\prime}
\newcommand{\cB}{{\cal B}}
\newcommand{\cO}{{\cal O}}
\newcommand{\vecx}{\vec{x}}
\newcommand{\vecy}{\vec{y}}
\newcommand{\vecp}{\vec{p}}
\newcommand{\vecq}{\vec{q}}
\newcommand{\tr}{{\rm tr} }
\newcommand{\Tr}{{\rm Tr} }
\newcommand{\NP}{Nucl. Phys. }

\newcommand{\cL}{{\cal L}}
\newcommand{\cA}{{\cal A}}
\newcommand{\cT}{{\cal T}}
\newcommand{\cD}{{\cal D}}
\newcommand{\cH}{{\cal H}}

\def\th{\theta}

\def\sst#1{{\scriptscriptstyle #1}}
\def\0{{\sst{(0)}}}
\def\1{{\sst{(1)}}}
\def\2{{\sst{(2)}}}
\def\3{{\sst{(3)}}}
\def\4{{\sst{(4)}}}
\def\5{{\sst{(5)}}}
\def\6{{\sst{(6)}}}
\def\7{{\sst{(7)}}}
\def\8{{\sst{(8)}}}
\def\9{{\sst{(9)}}}
\def\p{{\sst{(p)}}}
\def\q{{\sst{(q)}}}

\def\ssa{{\sst{(\alpha)}}}
\def\ssb{{\sst{(\beta)}}}
\def\ssg{{\sst{(\gamma)}}}
\def\j{{\sst{(j)}}}

\def\ve{\varepsilon}
\def\vf{\varphi}
\def\F{\Phi}
\def\wg{\wedge}

\def\thb{\bar{\theta}}
\def\Thb{\bar{\Theta}}
\def\barp{\bar{p}}
\def\barq{\bar{q}}
\def\barc{\bar{c}}
\def\bard{\bar{d}}
\def\e{\epsilon}

\def \bi{\bibitem}
\def \la {\label}

\def \l {\lambda}
\def\foot{\footnote}
\def \tl  {{\tilde \l}}
\def \sql {{\sqrt \l}}
\def \adss {$AdS_5 \times S^5$\ }
\newcommand{\rf}[1]{(\ref{#1})}
\def \ov {\over}

\def\Th{\Theta}
\def\vth{\vartheta}
\def\btheta{{\bar\theta}}
\def\ttheta{{{\tilde\theta}}}
\def\bttheta{{{\bar\ttheta}}}
\def\vth{\vartheta}

\def\ra{\rightarrow}
\def\N{{\cal N}}
\def\uM{\underline{M}}
\def\uA{\underline{A}}
\def\uN{\underline{N}}
\def\uP{\underline{P}}
\def\ua{\underline{a}}
\def\ub{\underline{b}}
\def\uc{\underline{c}}
\def\ud{\underline{d}}
\def\ue{\underline{e}}
\def\uf{\underline{f}}
\def\ui{\underline{i}}
\def\uj{\underline{j}}
\def\uk{\underline{k}}
\def\ual{\underline{\alpha}}
\def\ube{\underline{\beta}}
\def\um{\underline{m}}
\def\un{\underline{n}}
\def\up{\underline{p}}
\def\uq{\underline{q}}
\def\ur{\underline{r}}
\def\us{\underline{s}}

\def\umu{\underline{\mu}}
\def\unu{\underline{\nu}}
\def\ula{\underline{\l}}
\def\uka{\underline{\k}}
\def\usi{\underline{\s}}
\def\urh{\underline{\r}}
\def\cc{\circ}
\def\eqv{\equiv}

\def\ni{\noindent}

\def\Ep{E^{{}^{(+)}}}
\def\Em{E^{{}^{(-)}}}

\def\Mp{M^{{}^{(+)}}}
\def\Mm{M^{{}^{(-)}}}

\def \ha{{1\ov 2}}

\def\r{\rho}

\def\Y{{\rm Y}}
\def\X{{\rm X}}
\def\tY{\tilde{\rm Y}}
\def\tX{\tilde{\rm X}}
\def\dY{\dot{\rm Y}}
\def\dX{\dot{\rm X}}

\def \J {\mathcal{J}}
\def \del {\partial}

\def\dF{\dot{F}}
\def\dG{\dot{G}}
\def\df{\dot{f}}
\def\dx{\dot{x}}
\def \E {{\cal E}}
\def \S {{\cal S}}
\def \J {{\cal J}}

\def\ms{\mathcal{S}}
\def\mj{\mathcal{J}}
\def\soj{\fr{\ms}{\mj}}
\def \R {{\bf R}}
\def \om {\omega}
\def \bE {\bar E}
\def \x {{\cal X}}

\def \bi{\bibitem}
\def \la {\label}

\def \l {\lambda}
\def\foot{\footnote}
\def \tl  {{\tilde \l}}
\def \sql {{\sqrt \l}}
\def \adss {$AdS_5 \times S^5$\ }
\def \ov {\over}

\def \varpi {{\rm w}}

\def\thb{\bar{\theta}}
\def\Thb{\bar{\Theta}}
\def\mb{\bar{\m}}
\def\ab{\bar{\a}}
\def\zb{\bar{z}}
\def\psib{\bar{\psi}}
\def\barp{\bar{p}}
\def\barq{\bar{q}}
\def\barc{\bar{c}}
\def\bard{\bar{d}}
\def\e{\epsilon}
\def\wb{\bar{w}}
\def\lb{\bar{\l}}
\def\Jb{\bar{J}}
\def\Nb{\bar{N}}
\def\Zb{\bar{Z}}
\def\pab{\bar{\pa}}

\def\At{\tilde{A}}
\def\Bt{\tilde{B}}
\def\Ct{\tilde{C}}
\def\Dt{\tilde{D}}
\def\Et{\tilde{E}}
\def\Ft{\tilde{F}}
\def\Gt{\tilde{G}}
\def\Ht{\tilde{H}}
\def\Mt{\tilde{M}}
\def\Rt{\tilde{R}}
\def\at{\tilde{a}}
\def\bt{\tilde{b}}
\def\ct{\tilde{c}}
\def\dt{\tilde{d}}
\def\et{\tilde{e}}
\def\ft{\tilde{f}}
\def\gt{\tilde{g}}
\def\mt{\tilde{\mu}}
\def\nt{\tilde{\nu}}

\def\asth{\hat{*}}
\def\phh{\hat{\phi}}

\def\bA{{\bf A}}

\def\ola{\overleftarrow}
\def\ora{\overrightarrow}
\def\alt{\tilde{\a}}

\def\eh{\hat{e}}
\def\eph{\hat{\e}}
\def\ph{\hat{p}}
\def\alh{\hat{\a}}
\def\beh{\hat{\b}}
\def\gah{\hat{\g}}
\def\Fh{\hat{F}}
\def\muh{\hat{\m}}
\def\nuh{\hat{\n}}
\def\thh{\hat{\th}}
\def\dh{\hat{d}}
\def\ih{\hat{i}}
\def\jh{\hat{j}}
\def\kh{\hat{k}}
\def\deh{\hat{\d}}
\def\wh{\hat{w}}
\def\lah{\hat{\l}}
\def\Ah{\hat{A}}
\def\Ch{\hat{C}}
\def\Omh{\hat{\Omega}}

\def\xh{\hat{x}}

\def\ps{\rlap{\, /}\;\,p }
\def\ks{\rlap{\, /}\;\,k }

\def\gym{g_{YM}}

\def\adot{\dot{a}}
\def\bdot{\dot{b}}
\def\bpa{\bar{\pa}}

\def\pr{\prime}
\def\ssk{\medskip}
\def\bsk{\bigskip}

\def\clb{\color{blue}}
\def\clr{\color{red}}
\def\clv{\color{violet}}
\def\clg{\color{dolgreen}}
\def\clu{\color{UniBlue}}
\def\cly{\color{yellow}}

\def\t{\tau}
\def\cM{\mathcal{M}}
\def\S{\Sigma}

\def\N{\nabla}

\def\cR{\mathcal{R}}
\def\cL{\mathcal{L}}

\def\hb{\hbar}
\def\an{\hat{a}}
\def\ac{\hat{a}^\dag}
\def\hp{\hat{p}}

\def\Ec{{\cal E}}

\DeclareFontFamily{U}{FdSymbolA}{}
\DeclareFontShape{U}{FdSymbolA}{m}{n}{
    <-> s * [1] FdSymbolA-Book
}{}
\DeclareFontShape{U}{FdSymbolA}{m}{b}{
    <-> s * [1] FdSymbolA-Medium
}{}
\DeclareSymbolFont{fdsymbols}{U}{FdSymbolA}{m}{n}
\SetSymbolFont{fdsymbols}{bold}{U}{FdSymbolA}{m}{b}

\DeclareMathSymbol{\medtriangleright}{\mathbin}{fdsymbols}{86}
\DeclareMathSymbol{\medtriangleup}{\mathbin}{fdsymbols}{87}
\DeclareMathSymbol{\medtriangleleft}{\mathbin}{fdsymbols}{88}
\DeclareMathSymbol{\nabla}{\mathbin}{fdsymbols}{89}

\begin{document}

\title{
{\Large{\bf 
Taking the Null-Hypersurface Limit in the Parikh-Wilczek Membrane Approach
}}
}
{
\author{A.M. Arslanaliev$\,^{\vardiamondsuit,\spadesuit}$\footnote{arslanaliev.kh@gmail.com}
\,\, \,and  \,A.J. Nurmagambetov$\,^{\spadesuit,\vardiamondsuit,\varheartsuit}$\footnote{ajn@kipt.kharkov.ua; a.j.nurmagambetov@gmail.com}
\\ \\
$\,^{\vardiamondsuit}${ \it {\normalsize Department of Physics \& Technology, Karazin Kharkiv National University,}}\\
{ \it {\normalsize 4 Svobody Sq., Kharkiv 61022, Ukraine} }
\\
$\,^{\spadesuit}${ \it {\normalsize Akhiezer Institute for Theoretical Physics of NSC KIPT}}\\
{ \it {\normalsize 1 Akademichna St., Kharkiv 61108, Ukraine} }
\\
$\,^{\varheartsuit}${ \it {\normalsize Usikov Institute of Radiophysics and Electronics}}\\
{ \it {\normalsize 12 Ak. Proskury, Kharkiv 61085, Ukraine} }
}

\date{}

\maketitle

\vspace{-1cm}

\abstract{
We consider subtleties of the horizon (null-hypersurface) limit in the Parikh-Wilczek Membrane Approach to Black Holes. Specifically, we refine the correspondence between the projected Einstein equations of gravity with matter and the Raychaudhuri-Damour-Navier-Stokes (RDNS) equations of relativistic hydrodynamics. For a general configuration of gravity with matter we obtain additional terms in the hydrodynamic equations, which include very specific combinations of the contracted logarithmic derivatives of a parameter (the regularization function) determining the proximity of a stretched membrane to the black hole horizon. Nevertheless, direct computations of the new terms for exact (Schwarzschild and Kerr) black hole solutions prompt the standard form of the RDNS equations, due to the non-expanding horizon property of these solutions.   Therefore, the reduction of the extended RDNS equations to their classical form may be viewed as an additional consistency condition in the exact black hole solutions hydrodynamics, and may serve as a non-trivial test for various viable approximations of spacetime metrics. We compare in detail the Parikh-Wilczek Membrane Approach with the Gourgoulhon-Jaramillo method of a null-hypersurface description, as well as give the link of the obtained results to our previous work on the Kerr black holes.
}

\ssk
{{\bf Keywords}: Black Holes; Membrane Paradigm; Relativistic Hydrodynamics} 

\bsk
\newpage

\section{Introduction}

The Membrane Paradigm \cite{Thorne:1986iy} is one of the prominent ways to describe effective degrees of freedom on a Black Hole (BH) horizon. According to the Paradigm, a BH horizon is modeled by a stretched, penetrable and impacted by electromagnetic field membrane,  dynamics of which is given by hydrodynamic-type equations for a viscous relativistic fluid \cite{Damour:1978cg,Damour:1982,Parikh:1997ma}. In this way, the collective dynamics of fields near the event horizon is substituted by the dynamics of the dual fluid.

Interest in the hydrodynamic dual description of non-gravitational fields was increased after the AdS/CFT Duality foundation, and, as it was realized, some of the predictions of the Membrane Paradigm are directly related to outcomes of the AdS/CFT. Nevertheless, 
the Membrane Paradigm is in no way equated to the AdS/CFT correspondence \cite{Kovtun:2005ev,deBoer:2014xja}. Though a similarity between these approaches was mentioned since the early stages of the dual CFT hydrodynamics development \cite{Kovtun:2003wp,Iqbal:2008by}, mainly due to the universal character of the transport coefficients of the dual fluid \cite{Kovtun:2003wp,Iqbal:2008by,Kovtun:2008kx,Ritz:2010zza}, the Membrane Paradigm can, at best, be treated as a leading AdS/CFT approximation, or as its low-energy limit (see \cite{Bredberg:2010ky,Faulkner:2010jy} in this respect).  
Yet, further advances of the Membrane Paradigm may open new prospects in the AdS/CFT Duality progress. 

In our previous work, Ref. \cite{Nurmagambetov:2022wcw}, we extended the Membrane Paradigm to the case of rotating BHs.\footnote{Strictly speaking, in \cite{Nurmagambetov:2022wcw} we used the Membrane Paradigm in part, since we solely focused on the external part of the Kerr spacetime. We thank Prof. O.B. Zaslavskii for comments in this respect. Note, however, that this restriction is enough in solving for the problem how a black hole is viewed for an external observer as a “fluid”.} Operating with the Kerr solution in the Boyer-Lindquist coordinates, we
came to the conclusion on the divergence of the momentum density of the dual fluid on the horizon. In General Relativity the divergence of a quantity on the horizon may be caused by the coordinates choice.
So that one of the motivations for this paper is to re-derive the main characteristics of the dual fluid in the Eddington-Finkelstein parametrization of the Kerr metric, and to study their behavior in the vicinity of the horizon.

Accomplishing our goals requires the revision of main equations for the dual fluid, containing as the transport coefficients, as well as other basic characteristics -- energy, pressure, expansion, the momentum vector and the shear tensor -- of the medium. Previously, in \cite{Nurmagambetov:2022wcw}, we derived the transport coefficients etc. of the effective dual medium by comparing the energy-momentum tensor (EMT) of the stretched membrane with the conventional EMT of a relativistic viscous fluid. Here, we will recover the characteristics of the fluid from hydrodynamic-type equations, to which the projected, onto a null hypersurface, Einstein equations with matter are reduced. 

Specifically, 1+3 decomposition of time-like and space-like directions reduces the GR equations to external/internal geometry of a hypersurface, embedded into the target space. These equations are well-known as the Gauss and Codazzi-Mainardi equations (see, e.g.,\cite{Misner:1973prb,Gourgoulhon:2005ch,Gourgoulhon:2005ng,Straumann:2013spu}).\footnote{Following Misner, Thorne and Wheeler \cite{Misner:1973prb}, we will refer to these equations as the Gauss-Codazzi equations.} Further division of spatial directions \cite{Parikh:1997ma} makes it possible to present the projected, onto a 2D hypersurface,
Gauss-Codazzi equations as the Raychaudhuri and the Navier-Stokes type equations \cite{Raychaudhuri:1953yv,Damour:1982,Parikh:1997ma,Gourgoulhon:2005ch,Gourgoulhon:2005ng,Padmanabhan:2010rp,Li:2017ljz}. The system of these equations\footnote{Since the expansion and the shear tensor are also characteristics of the fluid, we refer to the Raychaudhuri equation as to a hydrodynamic-type equation.} determines the transport coefficients and other mentioned characteristics of the dual to the stretched membrane effective substance. But there is a subtlety, related to the fact, that the Gauss-Codazzi equations become the hydrodynamic-type equations only in the null-hypersurface limit. Taking this limit is a non-trivial task, that should be performed with additional care.

Indeed, there is the apparent conceptual difference in the geometric description of space-like (a stretched membrane type) and null (a BH horizon type) hypersurfaces, embedded into 4D space-time of the Minkowski signature. For a space-like hypersurface one needs two orthogonal to the hypersurface time-like and space-like vectors. These vectors can be represented in terms of two linearly independent null vectors, which makes the description more universal. The case of a null-hypersurface, the intrinsic metric of which degenerates, requires coincidence of two linear-independent null-vectors. Therefore, in the null-hypersurface limit, when the stretched membrane becomes the event horizon, it comes to be important to obtain null-vectors from the originally time- and space-like ones, and to make them equal on the null-surface in the last step.

On the way to this end, we want to revise, first, the procedure of getting the Raychaudhuri and the Damour-Navier-Stokes (RDNS) equations \cite{Damour:1982,Parikh:1997ma,Gourgoulhon:2005ch,Gourgoulhon:2005ng} from the projected Gauss-Codazzi equations in the Parikh-Wilczek Membrane Approach. The main revision concerns the way of taking the horizon (the null-hypersurface) limit, i.e., of transition to finite on the horizon quantities by the regularization. Details of this procedure can be found, e.g., in \cite{Parikh:1997ma,Nurmagambetov:2022wcw}.

Within the Membrane Approach of Ref. \cite{Parikh:1997ma}, the null-hypersurface limit is organized as setting the regularization factor (some coordinate function) to zero. The role of this function is to provide the finiteness of the divergent on the event horizon stress-energy tensor of a stretched membrane. On the other hand, this regularization factor can be viewed as a degree of proximity of the membrane to the true horizon. The outcome of taking the null-hypersurface limit in the Membrane Approach, without a reference to the specific type of space-time, consists in the extension of the RDNS-type equations by terms with the contracted logarithmic derivatives of the regularization factor. This result can be found in Section 2. In this section we also formulate two conditions on the regularization factor, called hereafter as the ``consistency conditions'', the fulfillment of which reduces the extended RDNS equations to their classical version \cite{Raychaudhuri:1953yv,Damour:1982}.\footnote{Note that the generalization of the Damour-Navier-Stokes equation in the vicinity of the horizon has been obtained in Ref. \cite{Gourgoulhon:2005ch}. In two added Appendices B and C we give the detailed comparison of the approach of \cite{Gourgoulhon:2005ch} and the Membrane Approach we follow in the paper, as well as we outline the correspondence of the obtained by us consistency conditions to the null-hypersurface limit of \cite{Gourgoulhon:2005ch}.} 

Second, we want to verify the validity of the obtained consistency conditions on two exact solutions to the Einstein equations. This part of our studies is presented in Section 3. Here we consider the Schwarzschild and the Kerr solutions in the Eddington-Finkelstein coordinates. The simplicity of the Schwarzschild metric does not allow us to fully evaluate possible limitations associated with the consistency conditions: they satisfy identically in the case. Performing the relevant computations for the Kerr solution is a less trivial task. However, our consistency conditions hold even in this case. Since the fulfillment of these conditions requires the tight coordination of different elements of a space-time metric, we can expect the same outcome for any exact BH solution to the Einstein equations, where such coordination takes place. 
The ground for such expectations is based on the ideology of the Membrane Approach to regularize the divergent on the horizon quantities, and to construct in this way the effective dynamical description of the horizon hypersurface in terms of the finite variables. Therefore, the established consistency conditions should hold for various viable approximations of spacetime metrics in the appropriate approximation order as well.
\footnote{Then, the viability of an approximate solution for the space-time metric can be evaluated comparing the approximation orders of the solution and of the consistency conditions: the approximation order of the consistency conditions has to be the same, or higher than that of the solution.}

Conclusions contain a summary of our findings, their applications and further developments.  Appendix A includes details on the surface gravity, computation of which is another non-trivial check of the consistency of a BH-type solution. Since our research focuses on the description of null-hypersurfaces, we pay special attention to comparing the approach used here with the approach of \cite{Gourgoulhon:2005ch,Gourgoulhon:2005ng}. In Appendix B we demonstrate the equivalence of the generalized Damour-Navier-Stokes equation of Ref. \cite{Gourgoulhon:2005ch} to the DNS-type equation from Section 2.1, obtained within the Membrane Approach \cite{Parikh:1997ma}. Appendix C contains the overview of taking the null-hypersurface limit of the generalized DNS equation from Ref. \cite{Gourgoulhon:2005ch}, that again leads us to the consistency conditions from Section 2.2. Recall, 
nothing special to recover the standard form of the DNS equation was required in Ref. \cite{Gourgoulhon:2005ch} upon taking the horizon limit. We briefly comment on the nature of discrepancies between two approaches in this respect.

We use the following notation throughout the paper. The 4D metric signature is chosen to be the mostly positive one. All indices (no matter what kind of, Latin or Greek) are supposed to be the indices of 4D target space. $g_{ab}$, $h_{ab}$ and $\g_{ab}$ are the 4D metric, 3D and 2D induced metric tensors, respectively. The induced metrics of low-dimensional spaces are used as projection operators.  Then, $\nabla_a$ symbol denotes 4D covariant (w.r.t. $g_{ab}$) derivative; $^{3}\cD_a$ and $^{2}\mathbf{D}_a$ are the covariant derivates w.r.t. 3D and 2D induced metrics. The explicit form of $^{3}\cD_a$ and $^{2}\mathbf{D}_a$ is given in the main text of the paper. In Appendices B and C we use the conventions of Refs. \cite{Gourgoulhon:2005ch,Gourgoulhon:2005ng} to simplify the comparison of different approaches to each other.


\section{Relativistic Hydrodynamics of the Membrane Approach }

\subsection{From the Gauss-Codazzi to the Damour-Navier-Stokes and the Raychaudhuri equations}

The starting point of our consideration (see \cite{Parikh:1997ma} for details) is the Gauss-Codazzi equation in the $1+3$ decomposition of metric: 
\be
^{3}\mathcal{D}^bt_{ab} = -h^c_aT_{cd}n^d.
\la{GC}
\ee
Here  $t_{ab}$ is the 3D stretched membrane energy-momentum tensor,
\be
t_{ab} = \frac{1}{8\pi}(Kh_{ab}-K_{ab});
\la{tabdef}
\ee
$T_{ab}$ denotes the energy-momentum tensor (EMT) of matter fields. The space-like unit vector $n^a$ ($n_a n^a=1$) is orthogonal to the hypersurface of the stretched membrane, endowed with the induced metric $h_{ab}=g_{ab}-n_a n_b$. The extrinsic curvature tensor of the membrane hypersurface is determined by
\be
K_{ab} = h^\alpha_a h^\beta_b \nabla_\beta n_\alpha;
\la{Kabdef}
\ee
$K=g^{ab}K_{ab}$ is its trace. The l.h.s. of eq. \rf{GC} also involves the 3D covariant derivative ${^3}\mathcal{D}_a$, whose action is specified by
\be
^{3}\mathcal{D}^c t_{ab}\equiv h^{c\gamma}{h_a}^{\alpha} {h_b}^{\beta} \nabla_{\gamma} t_{\alpha\beta} .
\la{cov3D}
\ee
To bridge the Membrane Approach \cite{Parikh:1997ma} to relativistic hydrodynamics, we have to introduce the time-like unit vector $u^a$ ($u^a u_a= -1$) and to form the 2D (Euclidean) induced metric $\g_{ab}= h_{ab} + u_a u_b$. Then, by means of $u_a$ and $\g_{ab}$, the membrane EMT $t_{ab}$ is equivalently presented as 
\be
\hat{t}_{ab} \equiv8\pi\,t_{ab} = \mathcal{E}u_a u_b+ \mathcal{P}\g_{ab} + q_a u_b + q_b u_a + \tau_{ab}.
\la{tabdecomp}
\ee
From the point of view of the original four-dimensional metric, eq. \rf{tabdecomp} requires the $1+1+2$ metric decomposition
\be
g_{ab} = -u_a u_b + n_a n_b + \g_{ab},
\ee
where, by construction,
\be
n^a h_{ab}=0, \; t_{ab}n^b = 0, \; u^a n_a = 0, \; \g_{ab}= h_{ab} + u_a u_b\;\g_{ab}n^b = \g_{ab}u^b = 0. 
\la{ungammadef}
\ee

The physical meaning of quantities on the r.h.s. of eq. \rf{tabdecomp} is easy to derive from the Eckart approach to relativistic irreversible thermodynamics (see, e.g., Section 7.8 of \cite{Alcubierre:2008} in this respect). The first two terms on the r.h.s. of \rf{tabdecomp} are treated as the energy density and the pressure; they form the EMT of an ideal fluid. A viscous fluid description requires adding the shear tensor $\tau_{ab}$, responsible for anisotropic stresses, the heat flow vector $q_a$, as well as adding an extra contribution to the pressure due to the fluid viscosity. The heat flow vector and the shear tensor are characterized, in particular, by 
\[
q^b u_b = 0,\quad \tau_{ab}u^a=0,\quad \tau_{ab}=\tau_{ba},\quad \mathrm{Tr} \,\tau_{ab}=0.
\]
However, it is convenient to consider the other set of common for the Membrane Approach variables
\be
\hat{\theta} = - \mathcal{E}, \; \hat{g} = \mathcal{P}-\frac{\hat{\theta}}{2},\; q_a= - \hat{\Omega}_a, \; \tau_{ab} = -\s_{ab}\,,
\la{thetagqtaurel}
\ee
in terms of which $\hat{t}_{ab}$ takes the following form:
\be
\hat{t}_{ab}=-\hat{\theta}u_au_b - \hat{\sigma}_{ab} + \left(\frac{\hat{\theta}}{2}+\hat{g}\right)\gamma_{ab}-\hat{\Omega}_au_b - \hat{\Omega}_bu_a\,.
\la{tabhat}
\ee
On account of the orthonormality/orthogonality conditions \rf{ungammadef}, for $\hat{\theta}$, $\hat{\Omega}_a$, $\hat{g}$ and $\hat{\sigma}_{ab}$ we get 
\[
\hat{\theta} = 	-\hat{t}_{ab}u^au^b, \;\; \hat{\Omega}_a = \hat{t}_{cb}u^b\g^{c}_a\,,\;\; \hat{g}=\frac12\left(\hat{t}_{ab}\g^{ab} + \hat{t}_{ab}u^au^b\right),\;\;
\]
\be
\hat{\sigma}_{ab}= -\left(\hat{t}_{cd}\g^{c}_a\g^{d}_b -\fr12 \g_{ab}\left(\hat{t}_{cd}\g^{cd} \right)\right).
\la{vrb}
\ee

We now have everything we need to write eq. \rf{GC} as equations of a 2D viscous fluid. We project eq. \rf{GC} onto the transverse and the longitudinal with respect to $u^a$ directions to this end:
\be
\g^c_a\,{^3}\mathcal{D}^b\hat{t}_{cb} = -8\pi\g^c_aT_{cd}n^d, \quad u^a\,{^3}\mathcal{D}^b\hat{t}_{ab} = -8\pi T_{cd}u^cn^d.
\la{GCsplit}
\ee
Inserting the membrane EMT \rf{tabhat} into the orthogonal to $u^a$ part of the Gauss-Codazzi equations, and taking into account the orthogonality of $u^a$ and $n^a$ to $h_{ab}$ and $\g_{ab}$, the orthogonality of $\hat{\W}_a$ to $u^a$ and $n^a$, and the definition of the Lie derivative along a vector field $\xi^a$, we arrive at
\be
\begin{split}
\g_{a}^b\pa_b\left(\frac{\hat{\theta}}{2} + \hat{g}\right) - {^2}\mathbf{D}^b\hat{\s}_{ab} + \left[\g_{ac}\left(\hat{g}-\frac{\hat{\theta}}{2}\right)-\hat{\s}_{ac}\right]u^b\nabla_bu^c &+8\pi \g^c_aT_{cd}n^d =\g^c_a\mathcal{L}_u\hat{\Omega}_c+\hat{\Omega}_a{^3}\mathcal{D}^bu_b \\
&
+\g^c_a\hat{\Omega}^b(\nabla_bu_c - \nabla_cu_b)\,. 
\end{split}
\la{GCa}
\ee
Here ${^2}\mathbf{D}_a$ is the 2D (contracted) covariant derivative  determined by 
\be
^2\mathbf{D}^b \hat{\s}_{ab}=\g^{\delta}_a\g^{\beta\rho}\nabla_\beta\hat{\s}_{\delta\rho}\,.
\la{cov2D}
\ee

For the part of eqs. \rf{GCsplit} along the time-like direction, after some algebra we get
\be
u^b\pa_b\hat{\theta}+\left(\frac{\hat{\theta}}{2}-\hat{g}\right)\,{^3}\mathcal{D}^bu_b + \hat{\s}_{ab}{^3}\mathcal{D}^bu^a + {^3}\mathcal{D}^b\hat{\Omega}_b + \hat{\Omega}_a u^b\nabla_bu^a+8\pi\, u^an^bT_{ab}=0. 
\la{Raych}
\ee
To outline the correspondence of eq. \rf{Raych} to the Raychaudhuri equation, it is convenient to introduce a symmetric tensor
\be
\hat{\Th}_{ab} = \hat{\s}_{ab} + \frac{\hat{\theta}}{2}\g_{ab},
\la{Thdef}
\ee
and to take into account that ${^3}\mathcal{D}^bu_b = h_{ab}{^3}\mathcal{D}^bu^a =  (\g_{ab}-u_au_b){^3}\mathcal{D}^bu^a = \g_{ab}\,{^3}\mathcal{D}^bu^a$. Then eq. \rf{Raych} becomes
\be
u^b\pa_b\hat{\theta}-\hat{g}\,{^3}\mathcal{D}^bu_b + \hat{\Th}_{ab}{^3}\mathcal{D}^bu^a + {^3}\mathcal{D}^b\hat{\Omega}_b + \hat{\Omega}_a u^b\nabla_bu^a+8\pi\, u^an^bT_{ab}=0. 
\la{GCb}
\ee

The so obtained eqs. \rf{GCa} and \rf{GCb} turn out to be the Damour-Navier-Stokes and Raychaudhuri equations of 2D relativistic hydrodynamics in the null-hypersurface (horizon) limit. Let us see how it happens.

\subsection{The Null-Hypersurface Limit }

General analysis of the stretched membrane EMT \rf{tabhat} leads to the conclusion on its divergence on the horizon \cite{Parikh:1997ma}. Geometrically, this fact is related to the degeneration of the stretched membrane hypersurface to null-hypersurface, that, in particular, means the divergence of time-like and space-like vectors $u^a$ and $n^a$ on the event horizon $\mathcal{H}$. The Membrane Approach \cite{Parikh:1997ma} suggests introducing a regularization factor (a function of coordinates) $\a$, which vanishes on the horizon, and whose role is to provide the finiteness of quantities in the null-hypersurface ($\a \ra 0$) limit. The choice of this regularization factor is determined by the requirements 
\be
\lim_{\a\rightarrow0}\a u^a=l^a, \quad   \lim_{\a\rightarrow0}\a n^a=l^a,\quad l^a l_a=0,
\la{rel1}
\ee
where $l^a$ is a null geodesic generator of $\mathcal{H}$. This null-vector obeys the equation
\be
l^b\nabla_b l^a = g_\mathcal{H}\,l^a, 
\la{leq}
\ee
which can be treated as a definition of
the surface gravity $g_\mathcal{H}$.\footnote{More on computations of the surface gravity for the Kerr BH can be found in Appendix A.} 

Now we have to regularize the EMT \rf{tabhat}, and to write down eqs. \rf{GCa} and \rf{GCb} in terms of the regularized (in the $\a\rightarrow0$ limit) quantities on the horizon. The regularization comes as follows:\footnote{The scaling in $\a$ of different variables depends on their physical interpretation. The H\'aji\v{c}ek field \cite{Hajicek:1974oua} $\hat{\W}_a$ is a measure of rotation, and it does not depend on a specific spacetime point, though it depends on the chosen frame.}
\be
\hat{\theta} = \a^{-1}\theta, \;\; \hat{g} = \a^{-1}g,\;\;\hat{\Theta}_{ab}=\a^{-1}\Theta_{ab},\;\;\hat{\sigma}_{ab} = \a^{-1}\sigma_{ab}, \;\; \hat{\Omega}_a = \Omega_a .
\la{reg}
\ee
Then, in terms of the regular on the horizon variables $\theta$, $g$, $\Th_{ab}$ and $\W_a$, eq. \rf{GCa} comes into
\be
\begin{split}
	&\a^{-1}\left(\g_{a}^b\pa_b\left(\frac{\theta}{2} + g\right) - {^2}\mathbf{D}^b\s_{ab} + \left[\g_{ac}\left(g-\frac{\theta}{2}\right)-\s_{ac}\right]u^b\nabla_bu^c-\left[\g_{ab}\left(\frac{\theta}{2} + g\right)-\s_{ab}\right]\pa^b\ln\a \right)\\
	&\hspace{2.7cm}+8\pi \g^c_aT_{cd}n^d =\g^c_a\mathcal{L}_u\Omega_c+\Omega_a{^3}\mathcal{D}^bu_b+\g^c_a\Omega^b(\nabla_bu_c - \nabla_cu_b)\,. 
\end{split}
\la{GCareg}
\ee
Whereas eq. \rf{GCb} becomes
\be
\a^{-1}\left(u^b\pa_b\theta - \th u^b\pa_b\ln\a - g\,{^3}\mathcal{D}^bu_b + \Th_{ab}{^3}\,\mathcal{D}^bu^a\right) +{^3}\cD^b \W_b + \Omega_a u^b\nabla_bu^a+8\pi\, u^an^bT_{ab} = 0\;.
\la{GCbreg}
\ee

The next step in completing the task is to take the limit $\a\rightarrow0$. Here we have to use both relations \rf{rel1} with care, since two operations -- taking the limit and acting by derivatives on $u^a$ and $n^a$ -- do not commute. And final expressions will be simplified by use of various orthogonality relations. 

Consider, for instance, two combinations with the 3D covariant derivative acting on $u^a$. The first combination, which occurs in both eqs. \rf{GCareg}, \rf{GCbreg}, is ${^3}\mathcal{D}^bu_b$. In the null-hypersurface limit we get
\[
\lim_{\a \ra 0} {^3}\mathcal{D}^bu_b = \lim_{\a \ra 0}\, h^{ab}\nabla_au_b = \lim_{\a \ra 0}\, \left(\g^{ab} - u^au^b\right)\nabla_a u_b =\lim_{\a \ra 0}\,\g^{ab}\nabla_au_b = -\lim_{\a \ra 0}\,u_b\nabla_a\g^{ab}  
\]
\be
\simeq - \a^{-1}l_b\nabla_a\g^{ab}=\a^{-1}\g^{ab}\nabla_a l_b = \a^{-1}\theta, \la{3Du1}
\ee
where we have used the orthogonality relations  $\g^{ab}u_b = 0$ and $\g^{ab}l_b = 0$ to rearrange the action of the derivative and to take the limit directly. To arrive at the final answer, we have used the definition of the expansion $\theta$ on the horizon, i.e., on the hypersurface, where relations \rf{rel1} hold: $\theta=\g^{ab}\nabla_al_b$.

The second combination, ${^3}\mathcal{D}^bu^a$, admits the following representation in the limit:
\be
\begin{split}
&\lim_{\a \ra 0} {^3}\mathcal{D}^bu^a =\lim_{\a \ra 0}  h^{b\beta}h^{a\a}\nabla_\beta u_\a = \lim_{\a \ra 0}  h^{b\beta}\g^{a\a}\nabla_\beta u_\a =-\lim_{\a \ra 0}  h^{b\beta}u_\a\nabla_\beta\g^{a\a} \\
&\simeq -\a^{-1}h^{b\beta}\,l_\a\nabla_\beta\g^{a\a} = \a^{-1}h^{b\beta}\g^{a\a}\nabla_\beta l_\a . 
\end{split}
\ee
It can be used to write down $\Th_{ab}{^3}\mathcal{D}^bu^a$ as $\a^{-1}\Th_{ab}\Th^{ba}$ on the horizon. Indeed, 
\be
\lim_{\a \ra 0} \Th_{ab}\,{^3}\mathcal{D}^bu^a\simeq \a^{-1}\Th_{ab}\,h^{b\beta}\g^{a\a}\nabla_\beta l_\a = \a^{-1}\Th_{ab}\,\g^{b\beta}\g^{a\a}\nabla_\beta l_\a = \a^{-1}\Th_{ab}\Th^{ba},
\ee
where we have applied the definition of $\Th_{ab}$ on the horizon: $\Th_{ab} = \g^{\a}_a\g^{\beta}_b\nabla_\a l_\beta$. Therefore, at this stage of our consideration, eq. \rf{GCbreg} turns into
\be
l^b\pa_b\theta - \theta\, l^b\pa_b\ln\a - g\theta + \Th_{ab}\,\Th^{ab} + 8\pi\, l^al^bT_{ab}+\lim_{\a \ra 0} \a^2\, \W_a\, u^b \nabla_b u^a= 0.
\la{GCbreg1}
\ee
And the non-triviality of the last term on the l.h.s. of \rf{GCbreg1} strongly depends on the scaling, with respect to the regularization factor $\a$, properties of $\lim_{\a \ra 0} u^b \nabla_b u^a$.

Let us consider this expression in more detail. Taking the $\a\ra 0$ limit of $v_a \equiv \g^{c}_a\,u^b\nabla_b u_c$, we get:
\be
\lim_{\a \ra 0} v_a \equiv \lim_{\a \ra 0} \g^{c}_a\,u^b\nabla_bu_c = - \lim_{\a \ra 0}u^bu_c\nabla_b\g^c_a \simeq -\a^{-2}l^bl_c\nabla_b\g^c_a=\a^{-2}\g^{c}_a\,l^b\nabla_b l_c\,.
\la{valimit}
\ee
Were we use eq. \rf{leq} as is, the introduced vector $v_a$ would be always equal to zero on the horizon, due to the orthogonality of the null-vector $l^a$ to the induced metric $\g_{ab}$. However, in the vicinity of the horizon, eq. \rf{leq} can be generalized to
\be
l^b\nabla_bl^c = g_\mathcal{H}l^c + \lambda^c,
\la{leqnew}
\ee
where $\lambda_c$ is a vector, which vanishes on the event horizon:
\be
\lim_{\a \ra 0} \lambda^a=0.
\la{lambdavec}
\ee
If $\lambda^a$ vanishes as $\a^2$ (e.g., $\lambda^c=\a^2 \g^{cd} \xi_d$), then $\lim_{\a \ra 0} v_a \ne 0$, so that $v_a$ remains finite on the horizon.\footnote{In Section 3 we justify the finiteness of $v_a$ in the horizon limit for the Kerr BH solution by direct computations.} Nevertheless, even with such a generalization, the last term on the l.h.s. of \rf{GCbreg1} becomes equal to zero.

To take the null-hypersurface limit of eq. \rf{GCareg}, one needs to write down the r.h.s. of this equation. Straightforward computations which take into account the orthogonality of $\W_a$ to $l^a$, symmetry of $\Th_{ab}$ tensor, and the outcome of eq. \rf{3Du1}, result in
\be
\lim_{\a \ra 0} \left[\g^c_a\mathcal{L}_u\Omega_c+\Omega_a{^3}\mathcal{D}^bu_b+\g^c_a\Omega^b(\nabla_bu_c - \nabla_cu_b)\right] \simeq \a^{-1}\left(\mathcal{L}_l\Omega_a + \Omega_a\theta\right) .
\ee
So that, the null-hypersurface limit of eq. \rf{GCareg} leads to
\be
\begin{split}
\g_{a}^b\pa_b\left(\frac{\theta}{2} + g\right) - {^2}\mathbf{D}^b\s_{ab} &+ \left[\g_{ac}\left(g-\frac{\theta}{2}\right)-\s_{ac}\right]v^c-\left[\g_{ab}\left(\frac{\theta}{2} + g\right)-\s_{ab}\right]\pa^b\ln\a\\
&+8\pi \g^c_aT_{cd}l^d = \g^c_a\mathcal{L}_l\Omega_c+\Omega_a\theta . 
\end{split}
\la{GCalim1}
\ee

Summing up, in the null-hypersurface limit the projected Gauss-Codazzi equations \rf{GCa} and \rf{GCb} are rearranged into
\be
l^b\pa_b\theta - g\theta + \Th_{ab}\,\Th^{ab} + 8\pi\, l^al^bT_{ab}= \theta\, l^b\pa_b\ln\a ,
\la{GCbreg2}
\ee
and
\be
\begin{split}
&\g_{a}^b\pa_b\left(\frac{\theta}{2} + g\right) - {^2}\mathbf{D}^b\s_{ab} +8\pi \g^c_aT_{cd}l^d-\g^c_a\mathcal{L}_l\Omega_c- \Omega_a\theta  \\
&= \left[\g_{ab}\left(\frac{\theta}{2} + g\right)-\s_{ab}\right]\pa^b\ln\a-\left[\g_{ac}\left(g-\frac{\theta}{2}\right)-\s_{ac}\right]v^c. 
\end{split}
\la{GCalim}
\ee
These equations coincide (cf., e.g., Ref. \cite{Gourgoulhon:2005ch,Gourgoulhon:2005ng}) with the Raychaudhuri and the Damour-Navier-Stokes (RDNS) equations, if the following conditions are satisfied:
\be
\left[\g_{ac}\left(g-\frac{\theta}{2}\right)-\s_{ac}\right]v^c=\left[\g_{ab}\left(\frac{\theta}{2} + g\right)-\s_{ab}\right]\pa^b\ln\a\,,\qquad l^b \pa_b \ln \a=0.
\la{rel2}
\ee
Since these conditions contain the potentially divergent on the horizon parts, while the proposed regularization procedure of \cite{Parikh:1997ma} was oriented toward making the quantities finite on the horizon, we arrive at the apparent contradiction. Therefore, we have to verify the fulfillment of these consistency conditions on a specific spacetime geometry. We will use two exact solutions to the Einstein equations~-- the Schwarzschild and the Kerr black holes -- to this end.

\section{Exploring the RDNS-type Equations of the Null-Hypersurface Limit }

\subsection{The Schwarzshild solution}

We get started with a warm-up exercise of the Schwarzschild solution, on the example of which we will establish/discuss: (i) the origin of different choices in the 1+1+2 metric decomposition within the Membrane Approach of \cite{Parikh:1997ma}; (ii) triviality of the consistency conditions \rf{rel2} for the Schwarzschild BH solution; (iii) the relation between the null-hypersurface limit of the Membrane Approach and the null-hypersurface description of \cite{Gourgoulhon:2005ch,Gourgoulhon:2005ng}.

To achieve our goals, we will use the Eddington-Finkelstein coordinates $(v,r,\theta,\vf)$, which are related to the original coordinates of the standard Schwarzschild metric $(t_S,r,\theta,\vf)$ as\footnote{We set $G=c=1$.}
\be
v=t_S+r^*=t_S+r+2M \ln \Big| \fr{r}{2M}-1 \Big|.
\la{v2tS}
\ee
The ``tortoise'' coordinate $r^*$ is the solution to the connection equation
\be
dr^*=\fr{dr}{f(r)},\qquad f(r)=1-\fr{2M}{r},
\la{tort}
\ee
and the Schwarzschild metric in the Eddington-Finkelstein coordinates becomes
\be
ds^2 = -f(r)dv^2 + 2dvdr + r^2(d\theta^2 + \sin^2\theta \,d\vf^2).
\la{SS}
\ee

To proceed further, we introduce a new time-like coordinate $t=v-r$, in terms of which the interval \rf{SS} turns into
\be
ds^2 = -\left(1-\frac{2M}{r}\right)dt^2 + \frac{4M}{r}dtdr + \left(1+\frac{2M}{r}\right)dr^2+r^2(d\theta^2 + \sin^2\theta\, d\vf^2).
\la{SS1}
\ee

Now, let's present the encoded in \rf{SS1} metric as $g_{ab}=-u_a u_b+ n_a n_b+\g_{ab}$. The structure of \rf{SS1} suggests two possible alternatives to this end:
\begin{itemize}
\item
First, we can choose 
\be
\begin{split}
	&{u}_a = \left(-\sqrt{1-\frac{2M}{r}},\frac{2M}{r\sqrt{1-\frac{2M}{r}}},0,0\right), \;\; {n}_a=\left(0,\frac{1}{\sqrt{1-\frac{2M}{r}}},0,0\right),\;\; -u^a u_a=n^an_a=1,
\end{split}
\la{tun}
\ee
\be
 {\g}_{ab} = 
 \begin{pmatrix}
 	0&&0&&0&&0\\
 	0&&0&&0&&0\\
 	0&&0&&r^2&&0\\
 	0&&0&&0&&r^2\sin^2\theta
 \end{pmatrix}. 
\la{tgamma}
 \ee
This choice corresponds to forming the perfect square from the 1st and the 2nd term on the r.h.s. of \rf{SS1}.
\item
Second, we can present the metric as $g_{ab}=-\tilde{u}_a \tilde{u}_b dx^a dx^b+\tilde{n}_a \tilde{n}_b dx^a dx^b+\g_{ab}dx^a dx^b$ with vectors
\be
\begin{split}
&\tilde{u}_a = \left(-\frac{1}{\sqrt{1+\frac{2M}{r}}},0,0,0\right), \;\; \tilde{n}_a= \left(\frac{2M}{r}\frac{1}{\sqrt{1+\frac{2M}{r}}},\sqrt{1+\frac{2M}{r}},0,0\right),\;\; -\tilde{u}^a \tilde{u}_a=\tilde{n}^a \tilde{n}_a=1,
\end{split}
\la{un}
\ee
and the same angle part ${\g}_{ab}$ as before. This presentation of the metric follows from forming the perfect square out of the 2nd and the 3rd term on the r.h.s. of eq. \rf{SS1}.
\end{itemize}
These two representations of the same metric are not unrelated to each other since the vectors are related by Lorentz transformations in the plane transversal to the angular coordinates:
\be
\tilde{u}_a={\Lambda_a}^b u_b,\qquad \tilde{n}_a={\Lambda_a}^b n_b;\qquad \Lambda \Lambda^T=1.
\la{Lorentzun}
\ee

Eqs. \rf{rel1}, crucial for the Membrane Approach, hold for the regularization factor\footnote{More on the choice of $\a$ can be found, e.g., in \cite{Parikh:1997ma,Nurmagambetov:2022wcw}.} $\a=\sqrt{f(r)}$,  and time/space-like vectors \rf{tun}:
\be
 \lim_{\a\rightarrow0}\a {u}^a = l^a, \;\; \lim_{\a\rightarrow0}\a {n}^a = l^a, \;\; l^a=(1,0,0,0).
\la{limAlpha20}
\ee
The vector $l^a$ is a null-vector on the horizon (i.e., at $f(r)=0$).\footnote{Outside (in the vicinity of) the horizon, $l^a$ becomes either a time-like vector, if it is associated with $u^a$, or a space-like one, if it is associated with $n^a$.} 
As one can see, the regularization factor depends only on the radial coordinate. Therefore, the consistency conditions \rf{rel2} are trivially satisfied, so that for the Schwarzschild geometry the RDNS-type equations \rf{GCbreg2}, \rf{GCalim}  coincide with that of originally derived in \cite{Damour:1982,Parikh:1997ma} and \cite{Raychaudhuri:1953yv}. 

Now, let us briefly discuss the correspondence of the Membrane Approach to the null-hypersurface description of \cite{Gourgoulhon:2005ch,Gourgoulhon:2005ng}. To define a time-like hypersurface, one can specify two null-vectors transversal/longitudinal to it. These null-vectors are constructed out of linear combinations of $\tilde{u}^a$ and $\tilde{n}^a$ (see \cite{Gourgoulhon:2005ng}),
\be
\begin{split}
&\hspace{2.5cm}
l^a=N(\tilde{u}^a+\tilde{n}^a),\qquad k^a=\fr1{2N}(\tilde{u}^a-\tilde{n}^a),\\
&
\tilde{u}^\a = \left(\sqrt{1+\frac{2M}{r}},-\frac{2M}{r}\frac{1}{\sqrt{1+\frac{2M}{r}}},0,0\right),\;\;\tilde{n}^a=\left(0,\frac{1}{\sqrt{1+\frac{2M}{r}}},0,0\right),
\end{split}
\la{lkdef}
\ee
with a lapse function $N$. To equate $l^a$ of \rf{lkdef} to $l^a=(1,0,0,0)$ on the event horizon $r_\cH=2M$, one fixes $N=1/\sqrt{1+2M/r}$. Then, after recovering the exact form of the second null-vector $k^a$, it is easy to verify that
$l^2=0$, $k^2=0$ and $l^a k_a=-1$ everywhere. 

Apparently, the same consideration is applicable to ${u}^a$ and ${n}^a$ vectors 
\be
{u}^a = \left(\frac{1}{\sqrt{1-\frac{2M}{r}}},0,0,0\right), \;\; {n}^a=\left(\frac{2M}{r\sqrt{1-\frac{2M}{r}}},\sqrt{1-\frac{2M}{r}},0,0\right),
\la{unup}
\ee
which are the contravariant counterpart of \rf{tun}. In this case, the lapse function $N$ is given by $N=\sqrt{(1-2M/r)}/(1+2M/r)$. Therefore, following \cite{Gourgoulhon:2005ch,Gourgoulhon:2005ng}, one may recover the corresponding null-vectors for any reasonable form of 1+1+2 metric decomposition. However, to describe a null-hypersurface the same conditions must be met as in eqs. \rf{limAlpha20}.  

To sum up, different rearrangements of the diagonal and non-diagonal terms in the non-angular part of metric \rf{SS1} lead to different forms of its 1+1+2 decomposition. Just one of them falls into the criteria of the null-hypersurface description, and can be used in computing characteristics of the dual, to the stretched membrane near the BH horizon, fluid. There are various approaches to reach this goal, examples of which are that of \cite{Parikh:1997ma} and \cite{Gourgoulhon:2005ch,Gourgoulhon:2005ng}. They are slightly different in details, but comparing them to each other\footnote{We refer the reader for two Appendices B and C, where we establish the equivalence between the generalized DNS equations near the event horizon of this paper and of Ref. \cite{Gourgoulhon:2005ch}, and re-derive the established here consistency conditions  from the construction of \cite{Gourgoulhon:2005ch,Gourgoulhon:2005ng}.} we draw the conclusion that they lead to the same outcomes. 

Unfortunately, the Schwarzschild solution is plain to reveal all sides of the RDNS equations extension. It can be done in the analysis of a more complicated example, like the Kerr BH solution, to the consideration of which we now turn.

\subsection{The Kerr Black Hole}

The Kerr metric in the Eddington-Finkelstein coordinates $(v,r,\theta,\vf)$ is given by 
\be
\begin{split}
ds^2 = &-\left(1-\frac{2Mr}{\rho^2}\right)dv^2 +2dvdr - 2a\sin^2\theta d\vf dr-\frac{4aMr}{\rho^2}\sin^2\theta dvd\vf  + \rho^2 d\theta^2\\
&\hspace{1.2cm}+\left(r^2+a^2+\frac{2Mr}{\rho^2}a^2\sin^2\theta\right) \sin^2\theta  d\vf^2,
\qquad \rho^2=r^2+a^2 \cos^2\theta.
\la{Kerrv}
\end{split}
\ee
As in the Schwarzschild BH case, we introduce the time coordinate $t=v-r$, so that, in terms of $(t,r,\theta,\vf)$, 
\be
\begin{split}
	&\hspace{1.2cm}ds^2 = -\left(1-\frac{2Mr}{\rho^2}\right)dt^2 +\frac{4Mr}{\rho^2}dtdr - \frac{4aMr}{\rho^2}\sin^2\theta dt d\vf +\left(1+\frac{2Mr}{\rho^2}\right)dr^2  \\
	&\hspace{10mm} -2a\sin^2\theta\left(1+\frac{2Mr}{\rho^2}\right) drd\vf + \rho^2d\theta^2 +\left(r^2+a^2+\frac{2Mr}{\rho^2}a^2\sin^2\theta\right) \sin^2\theta  \,d\vf^2 .
	\la{Kerrt}
\end{split}
\ee

The metric \rf{Kerrt} contains three cross-terms, that apparently complicates the $1+1+2$ decomposition. Its inverse contains merely two cross-terms, 
\be
d\tilde{s}^2\equiv g^{ab}\pa_a \pa_b=-\left(1+\fr{2Mr}{\rho^2}\right) \pa_t^2+\fr{4Mr}{\rho^2} \pa_t \pa_r+\fr{\D}{\rho^2} \pa^2_r+\fr{2a}{\rho^2}\pa_r \pa_\vf+\fr1{\rho^2}\pa_\theta^2+\fr{1}{\rho^2 \sin^2\theta} \pa^2_\vf ,
\la{dsKerrinv}
\ee
that slightly simplifies the computations. In writing the inverse metric we have used the dual basis notation
\be
dx^b \pa_a=\delta_a^b .
\la{derdif}
\ee
$\D$ is the standard for the Kerr solution function of the radial direction,
\be
\D=r^2+a^2-2Mr,
\la{Ddef}
\ee
used for determining the radial locations ($r^{\pm}_\cH$) of the black hole horizons: $\D(r^{\pm}_\cH)=0$.

As in the case of the Schwarzschild spacetime, there are two possible rearrangements of the inverse Kerr metric  \rf{dsKerrinv} as $g^{ab}=-u^a u^b+n^a n^b+\g^{ab}$ suggested by its structure: 
\begin{enumerate}[label=\roman*)]
\item
The first option refers to forming the perfect square from the 1st and the 2nd terms on the r.h.s. of \rf{dsKerrinv} at the first step, and going along this line further on.
\item
The second option supposes combining the 2nd and the 3rd terms on the r.h.s. of \rf{dsKerrinv} at the first stage, with developing this line after.\footnote{In the Schwarzschild (i.e., zero-rotation) limit, the option ``i)'' corresponds to $\tilde{u}^a$ and $\tilde{n}^a$ of \rf{lkdef}, while the option ``ii)'' leads to \rf{unup}.} 
\end{enumerate} 
However, within the Membrane Approach, we have to choose the way, along which we will be able to produce eqs. \rf{rel1} with the appropriately chosen $\a$. Thus, we have to determine the null-vector $l^a$ for the Kerr geometry first.

According to the Kerr metric structure, there are two associated Killing vectors (in $t$ and $\vf$ directions), that specifies non-trivial components of the null-vector $l^a$:
\be
l^a=(l^t,0,0,l^\vf)=(1,0,0,X).
\la{ldef}
\ee
The function $X$ is fixed from the null-vector condition, $l^a l_a=0$ . For the metric \rf{Kerrt}, the null-vector condition leads to
\be
X=\fr{2a Mr}{A}\pm \fr{\rho\sin\theta}{A}\sqrt{\left(A-(2Mr)^2\right)-2Mr \D}\,,
\la{Xlnul}
\ee
where we have introduced 
\be
A=\rho^2(r^2+a^2)+2a^2 Mr \sin^2\theta=(\rho^2+2Mr)\D+(2Mr)^2 .
\la{Adef}
\ee
On the horizon, where $\D=0$ and $A=(2Mr_\cH)^2$, eq. \rf{Xlnul} turns into $X=a/(2M r_\cH)$; hence\footnote{Note that here we consider the external part of the Kerr spacetime. Therefore, $r_\cH =M+\sqrt{M^2-a^2}$ is the outer horizon (the largest root of $\D(r)=0$ algebraic equation).}
\be
l^a=\left(1,0,0,\fr{a}{2Mr_\cH}\right).
\la{lnulhor}
\ee

It is easy to check that the metric decomposition ``i)'' does not lead to $\lim_{\a \ra 0} \a u^a=\lim_{\a \ra 0} \a n^a=l^a$, required in the Membrane Approach, whatever the $\a$ factor would be. For this reason, we have to turn to the option ``ii)''. Rearranging the metric \rf{dsKerrinv} in this way, we arrive at
\be
d\tilde{s}^2=\fr1{\rho^2} \left[ -\left(\sqrt{\fr{A}{\D}} \,\pa_t+\fr{2Mra}{\sqrt{A\D}}\,\pa_\vf \right)^2+\left(\fr{2Mr}{\sqrt{\D}}\,\pa_t+\sqrt{\D}\,\pa_r+\fr{a}{\sqrt{\D}}\,\pa_\vf \right)^2+\pa^2_\theta+\fr{\rho^4}{A \sin^2\theta}\,\pa^2_\vf \right].
\la{dsKerrinv1}
\ee
Now, for getting $g^{ab}=-u^a u^b+n^a n^b+\g^{ab}$, with $\D$ of \rf{Ddef} and $A$ of \rf{Adef}, we take
\be
u^a =\D^{-1/2}\left(\frac{\sqrt{A}}{\rho},0,0,\frac{2Mra}{\rho\sqrt{A}}\right) , \qquad n^a =\D^{-1/2}\left(\frac{2Mr}{\rho},\frac{\D}{\rho},0,\frac{a}{\rho}\right),
\la{unupKerr}
\ee
\be
\g^{ab} = \begin{pmatrix}
		0&&0&&0&&0\\
		0 && 0&&0&&0\\
		0&&0&&\rho^{-2}&&0\\
		0 &&0&&0&&\frac{\rho^2}{A\sin^2\theta} 
	\end{pmatrix} .
\la{gammainvKerr}
\ee
And, to equate \rf{lnulhor} and \rf{unupKerr} in the null-hypersurface limit, the regularization function has to be   
\be
\a = \rho\sqrt{\fr{\D}{A}} .
\la{alphaKerr}
\ee

Having fixed all the needed ingredients, we can compute the energy-momentum tensor $\hat{t}^{ab}$ (cf. eqs. \rf{tabdef}, \rf{Kabdef}, \rf{tabdecomp} and \rf{tabhat}): 
\be
\hat{t}^{ab} = \frac{1}{\D^{1/2}\rho^3(r,\theta)}\begin{pmatrix}
	-(r-M)a^2\cos^2\theta - a^2(M+r)-2r^3&&0&&0&&-aM\\
	0&&0&&0&&0\\
	0&&0&&r-M&&0\\
	-aM&&0&&0&&\frac{r-M}{\sin^2\theta}
\end{pmatrix} .
\ee
After that, taking into account eqs. \rf{unupKerr}, \rf{gammainvKerr}, and \rf{vrb}, we get 
\begin{equation}
\begin{split}
&\hat{\theta} = \frac{\Delta^{1/2}}{A\rho}\,h(r,\theta), \quad h(r,\theta) = 2r^3 + a^2(r+M) + (r-M)a^2\cos^2\theta; \\
&\hat{\Omega}^a = \left(0,0,0,\frac{aM}{\rho^2A^{3/2}}\,\omega(r,\theta)\right),\quad \omega(r,\theta) = a^2(a^2-r^2)\cos^2\theta - 3r^4 -a^2r^2; \\
&\hat{g} = \frac{M}{\Delta^{1/2}\rho^3A}\left((r^2+a^2)^2(r^2-a^2\cos^2\theta)-4a^2Mr^3\sin^2\theta\right);\\
&\hat{\sigma}^{ab}=\mathrm{diag}\left(0,0,\frac{\Delta^{1/2}}{\rho^3}\left(\frac{r}{\rho^2}-\frac{h(r,\theta)}{2A}\right),\frac{\Delta^{1/2}a^2}{2\rho A^2}\left(a^2(M-r)\cos^2\theta - r^2(3M+r)\right)\right).
\end{split}
\la{hatvarKerr}
\end{equation}
Recall, $\D$, $A$ and $\rho$ have been introduced in \rf{Kerrv}, \rf{Ddef} and \rf{Adef}. The tensor $\hat{\Th}_{ab}$, introduced in \rf{Thdef}, then becomes
\be
\begin{split}
	&\hat{\Th}^{ab}=\mathrm{diag}\left(0,0,\frac{\Delta^{1/2}r}{\rho^5},\frac{\Delta^{1/2}}{\rho\sin^2\theta}\frac{\rho^2 h(r,\theta)-Ar}{A^2}\right).
\end{split}
\la{ThKerr}
\ee

A brief inspection of \rf{hatvarKerr} and \rf{ThKerr} leads to the conclusion that the only $\hat{g}$ turns out to be singular on the horizon. After the regularization, $g = \a\hat{g}$ becomes
\be
g = \frac{M}{\rho^2A^{3/2}}\left((r^2+a^2)^2(r^2-a^2\cos^2\theta)-4a^2Mr^3\sin^2\theta\right),
\ee 
and, in the null-hypersurface limit, it coincides with the surface gravity on the horizon: 
\be
\lim_{\a \ra 0} g=g_\cH ,\qquad 
g_\mathcal{H} = \frac{\sqrt{M^2-a^2}}{2M(M + \sqrt{M^2-a^2})}. 
\la{gH}
\ee
The other non-trivial quantity on the horizon is the vector field $\hat{\W}_a$, which, according to \rf{reg}, does not need to be regularized: $\hat{\W}_a={\W}_a$. 

To figure out the form of the RDNS-type equations in the case, we have to verify the consistency conditions \rf{rel2}. For the null-vector $l^a$ (see eq. \rf{lnulhor}) and the regularization function $\a$ (see eq. \rf{alphaKerr}), the second condition of \rf{rel2} is satisfied. Verifying the first condition of \rf{rel2}, one needs the exact form of the projected acceleration vector $v_a = \g^c_a\,u^b\nabla_b u_c$, a direct computation of which results in
\be
v_a = \left(0,0,-\frac{2a^2Mr(r^2+a^2)\cos\theta\sin\theta}{\rho^2 A} ,0\right).
\la{vadef}
\ee
According to \rf{vadef}, $v_a$ is finite on the horizon, that has been assumed upon the derivation of the consistency conditions \rf{rel2}. Straightforward computations show that, with $v_a$ from \rf{vadef} and $\a$ from \rf{alphaKerr}, the first of the conditions \rf{rel2} is also satisfied. 

Thus, as in the Schwarzschild case, the Kerr BH geometry keeps the standard form (cf., e.g., \cite{Gourgoulhon:2005ng}) of the Raychaudhuri and the Damour equations of a (1+2) null-hypersurface. (I.e., eqs. \rf{GCbreg2} and \rf{GCalim} have trivial right hand sides in the case.)

We end up this section with recalling how the l.h.s. of the Damour equation \rf{GCalim} is related to the Navier-Stokes equation for a viscous fluid \cite{Damour:1982}. Let us introduce a force surface density $f_a=-\g_a^c T_{cd}l^d$, the momentum density $\pi_a$, the pressure $p$, the shear and bulk viscosities $\eta$ and $\zeta$ of the fluid as
\be
\pi_a=-\fr{1}{8\pi} \W_a,\quad p=\fr{g}{8\pi},\quad \eta=\fr{1}{16\pi},\quad \zeta=-\fr1{16\pi}.
\la{pietcdef}
\ee
Then, the l.h.s. of \rf{GCalim} can be presented in the form of the Navier-Stokes equation 
\be
\g_a^c \cL_l \pi_c+\theta \pi_a=-{}^2{\bf D}_a p+2 \eta\, {}^2{\bf D}^b \s_{ba}+\zeta \, {}^2{\bf D}_a \theta+f_a .
\la{NS}
\ee
The correspondence of the momentum density $\pi^a$ to the H\'aji\v{c}ek field $\W^a=\hat{\W}^a$ makes the former finite on the horizon. (Cf. eqs. \rf{hatvarKerr}). Comparing this result with the early obtained divergence of $\pi^a$ on the horizon of the Kerr BH in the Boyer-Lindquist coordinates \cite{Nurmagambetov:2022wcw}, we conclude on the frame dependence of the momentum density: the correct choice of coordinates makes the quantity finite on the horizon.

\section{Conclusions}

Let us summarize our findings. At the first stage of our studies, following the Parikh-Wilczek Membrane Approach to black holes, we have presented the Gauss-Codazzi equations on the horizon as hydrodynamic-type equations. We expected to derive the standard Raychaudhuri and the Damour-Navier-Stokes (RDNS) equations of a viscous fluid in this way. However, our actual result looks slightly different: the final equations are extensions of the RDNS equations. Specifically, there appears new terms, containing derivatives of a function of the regularization parameter. Recall, this parameter is used for making the energy-momentum tensor of a stretched membrane finite on the horizon. The explicit form of this function -- the logarithm in the case of the standard regularization within the Membrane Approach -- depends on the way of regularization. Anyway, the established new terms can not be ignored in the null-hypersurface limit, upon building the bridge between geometry (the Gauss-Codazzi equations) and dynamics (the RDNS equations).
Getting the ``classical'' RDNS equations back two non-trivial conditions must be met. And the fulfillment of these consistency conditions requires the tight coordination of different elements (metric, null-vectors, projected acceleration vector, regularization function) of the chosen space-time geometry.

To investigate this issue in more detail, we have examined two notable examples of exact solutions to the Einstein equations: the Schwarzschild and the Kerr black holes. The case of the Schwarzschild solution has been considered as a warm-up exercise, aimed at establishing the machinery, which could be further applied to the Kerr solution in the Eddington-Finkelstein parametrization. In view of simplicity of the Schwarzschild solution, the mentioned consistency conditions are trivially satisfied. The established consistency conditions have been verified, to the full extent, in the case of the Kerr metric in the Eddington-Finkelstein parametrization. 
The verification requires more technical efforts, due to a complicated structure of the metric tensor, but we arrive at the conclusion on the fulfillment of the consistency conditions in this case.
Therefore, for the Schwarzschild and the Kerr solutions, the RDNS equations of the Membrane Approach do not change. We can expect the same effect for exact BH solutions to the Einstein equations with the required tight coordination of the spacetime geometry components which results in the non-expanding (isolated) horizon \cite{Hajicek:1973,Hajicek:1974,Hajicek:1975,Ashtekar:1998sp}, and, consequently, in the classical form of the RDNS equations on the horizon. As we have mentioned in Introduction, the ground for such expectations is based on the ideology of the Membrane Approach to regularize the divergent on the horizon quantities, and to construct in this way the effective dynamical description of the horizon hypersurface in terms of the finite variables. 

In the course of our studies we paid a special attention to
the relation of the Membrane Approach \cite{Parikh:1997ma} to the Gourgoulhon-Jaramillo \cite{Gourgoulhon:2005ch,Gourgoulhon:2005ng} method of a null-hypersurface description. 
Note that within the approach of \cite{Gourgoulhon:2005ch,Gourgoulhon:2005ng} it was obtained the generalization of the Damour-Navier-Stokes (DNS) equation in the vicinity of the event horizon of a BH-type solution to the Einstein equations. However, in the null-hypersurface limit, the authors of \cite{Gourgoulhon:2005ch,Gourgoulhon:2005ng} drew the conclusion that on the horizon the generalized DNS equation is reduced to its classical form. This fact motivates us to investigate the correspondence between the Membrane Approach used in the paper and the approach of \cite{Gourgoulhon:2005ch,Gourgoulhon:2005ng} in more detail. In two added Appendices we establish the equivalence of the DNS equation generalizations near the horizon in the approaches of \cite{Parikh:1997ma} and 
\cite{Gourgoulhon:2005ch,Gourgoulhon:2005ng} for the specific metric parametrization used by Gourgoulhon and Jaramillo. (This parametrization supposes the trivial vorticity tensor of a geodesic congruence.) Also, the detailed consideration of the null-hypersurface limit within the approach of \cite{Gourgoulhon:2005ch,Gourgoulhon:2005ng} leads to the same set of the consistency conditions as in the Membrane Approach. It turns out that both consistency conditions follow from the generalized DNS equation without the need to consider an additional equation like the Raychaudhuri equation. This emphasizes the self-consistency of the Einstein equations, from which the RDNS equations follow. And again, the main condition to keep the classical form of the DNS equations on the horizon is the requirement of having the non-expanding horizon. 

Since the non-expanding/isolated horizon is common for exact BH solutions \cite{Hajicek:1973,Hajicek:1974,Hajicek:1975,Ashtekar:1998sp}, a more interesting situation arises for non-exact solutions of the BH type, like, for instance, slowly rotating BHs, metrics mimicking black holes, post-Newtonian corrected BHs etc., examples of which can be found in \cite{Hartle:1968si,Johannsen:2011dh,Rezzolla:2014mua,Konoplya:2016jvv,Konoplya:2023owh,Nashed:2023cyr,Hartong:2023yxo}. If these approximations of the spacetime metric are used in the construction of the relativistic hydrodynamics within the Membrane Approach, the established consistency conditions should hold for them as well, to the same order of the approximation as for the Einstein equations. (Cf. footnote 5 in this respect.) So that, the established here consistency conditions can be served as an additional tool in verifying the viability of such approximations. It would be interesting to find examples of metrics where the consistency conditions fails, and to analyze reasons for that. We hope to report on this and other results of our studies in future publications.

\bsk
{\bf Acknowledgements}. AJN is thankful to Prof. O.B. Zaslavskii for correspondence and viable comments. The work of A.J.N. is supported in part within the Cambridge-NRFU 2022 initiative ''Individual research (developments) grants for researchers in Ukraine (supported by the University of Cambridge, UK)'', project №2022.02/0052.

\ssk
{\bf Conflicts of interest}.The authors declare no conflict of interest.  


\bsk\bsk
\appendix
\numberwithin{equation}{section}

\section{The surface gravity for the Kerr solution}

Let us consider the surface gravity calculation for a rotating BH in more detail. 

There are several ways to compute the surface gravity. One may use, for instance, eq. \rf{leq}. Another way is to take into account the fact that the null-vector $l^a$ \rf{lnulhor} is nothing but the Killing vector of the Kerr BH metric, $\xi^a=\pa_t+\W \pa_\vf$, on the horizon. $\W$ is the angular velocity, defined by
\be
\W\equiv \fr{d\vf}{dt}=\fr{d\vf/ds}{dt/ds}=\fr{u^\vf}{u^t}=\fr{2Mra}{A} ,
\la{Omdef}
\ee
where we have used the components of $u^a$ from \rf{unupKerr}. Note, preliminarily, that \rf{Omdef} points to the following details: 
\begin{enumerate}[label=\roman*)]
\item
the vector $u^a$ is the velocity of the so-called {\it stationary} observer, which possesses arbitrary, but uniform, angular velocity $\W$;
\item
this angular velocity coincides with the ZAMO (zero angular momentum observer) angular velocity, defined by $\hat{L}\equiv u_\vf \xi^\vf=0$. For metric \rf{Kerrt}, the zero angular momentum is realized as $u_\vf=u^t g_{t\vf}+u^\vf g_{\vf\vf}=0$. Hence,
\be
\w\equiv \fr{u^\vf}{u^t}=-\fr{g_{t\vf}}{g_{\vf\vf}}.
\la{wZAMOdef}
\ee
With metric \rf{Kerrt},
\be
\w=\fr{2aMr}{A},
\la{wKerrt}
\ee
that is the same as $\W$ of \rf{Omdef}. Therefore, the 4-velocity $u^a$ from \rf{unupKerr} is that of a ZAMO;
\item
the angular velocity $\W$ coincides with the angular velocity of the black hole on the black hole horizon 
\be
\W_\cH\equiv \w(r_\cH)=\fr{a}{2M r_{\cH}}\equiv \fr{a}{r^2_\cH+a^2}\,.
\la{Omhordef}
\ee
\end{enumerate}

By use of the Killing vector nature of $\xi^a$, one may easy verify the relation
\be
\nabla_a \left(\xi^b \xi_b \right)=-2 g_\cH \,\xi_a,\qquad \xi^a=\pa_t+\W_\cH \pa_\vf ,
\la{surfgdef2}
\ee 
which we will use in computations of the surface gravity $g_\cH$.

The norm of $\xi^a$ for the Kerr metric in the Eddington-Finkelstein coordinates is given by
\be
\xi^a \xi_a=\fr{A\sin^2\theta}{\rho^2} \left(\W_\cH-\w \right)^2-\fr{\rho^2 \D}{A} .
\la{xinorm}
\ee
Then the covariant derivative of $\xi^a \xi_a$ on the horizon, where $\W_\cH=\w(r_\cH)$, is equal to
\be
\nabla_a  \left(\xi^b \xi_b \right)=-\fr{\rho^2}{A}\Big|_{\cH}\,\pa_a \D.
\la{derxinorm}
\ee
Or, with $\D=r^2+a^2-2Mr$, eq. \rf{derxinorm} becomes
\be
\nabla_a  \left(\xi^b \xi_b\right)=-\fr{2\rho^2}{A}\,(r-M)\Big|_\cH \pa_a r.
\la{derxinorm1}
\ee

Now, we compare the r.h.s. of \rf{derxinorm1} with $\xi_a$, and take both quantities on the horizon. We get
\be
\xi_a\Big|_\cH=\lim_{\a \ra 0} \a u_a\Big|_\cH=\left(0,\fr{2M r \rho^2}{A}\Big|_\cH,0,0\right),
\la{xihor}
\ee
so that, combining \rf{surfgdef2}, \rf{derxinorm} and \rf{xihor}, and taking all of these quantities on the horizon, we arrive at
\[
\fr{2\rho^2}{A}\left(r-M\right)\Big|_\cH=g_\cH \fr{2Mr \rho^2}{A}\Big|_\cH.
\]
Therefore,
\be
g_\cH=\fr{r_\cH-M}{2M r_\cH}.
\la{gHKerr}
\ee
Since $r_\cH=M+\sqrt{M^2-a^2}$, we recover eq. \rf{gH}. Equivalently, the surface gravity can be presented as
\be
g_\cH=\fr{r_+-r_-}{2(r^2_+ +a^2)},\qquad r_{\pm}=M\pm\sqrt{M^2-a^2}.
\la{gHKerr1}
\ee

\bsk
\section{On the Gourgoulhon's generalization of the DNS equation}

In this Appendix we provide the link between the RDNS equations \rf{GCbreg2}, \rf{GCalim} and the generalization of the DNS equation in the vicinity of a BH horizon, derived in Ref. \cite{Gourgoulhon:2005ch}. 

Let's start with an overview of basics in the construction of \cite{Gourgoulhon:2005ch}. Suppose we are dealing with a hypersurface $\tilde{{\cal H}}$, which is foliated by a family of 2D space-like surfaces. The orthogonal to these 2d surfaces plane can be generated by basic vectors $(\mathbf{h},\mathbf{m})$, one of which (say, the vector $\mathbf{h}$) is inside of $\tilde{\cal H}$, and the other one is orthogonal to $\tilde{\cal H}$. We refer the reader to Ref. \cite{Gourgoulhon:2005ch} for more details on the basic vectors $(\mathbf{h},\mathbf{m})$. For our purposes it would be enough to use the representation of these vectors in terms of null vectors $(\mathbf{l},\mathbf{k})$ on the whole 4D space-time:
\be
h^a = l^a - Ck^a,\qquad m^a = l^a + Ck^a.
\la{link}
\ee
From the properties of $(\mathbf{l},\mathbf{k})$, $l^2=0$, $k^2=0$ and $l^a k_a = -1$ it follows that
\be
C = \frac12 h^a h_a = -\frac12 m^a m_a .
\la{Cdef}
\ee
Since we are interested in a time-like hypersurface $\tilde{\cal H}$, $C<0$. 

In \cite{Gourgoulhon:2005ch} it was established the following generalization of the DNS equation on the hypersurface $\tilde{\cal H}$:
\be
q^c_a\mathcal{L}_h\Omega^{(\mathbf{l})}_c + \th^{(\mathbf{h})}\Omega_a^{(\mathbf{l})} = {^2}D_a\langle \kappa^{(\mathbf{l})}, \mathbf{h}\rangle - {^2}D^b\s^{(\mathbf{m})}_{ba} + \frac12\,{^2}D_a\th^{(\mathbf{m})}-\th^{(\mathbf{k})}\,{^2}D_aC+8\pi q^c_aT_{cd}m^d,
\la{gDNSb}
\ee
where $q_{ab}$ is the 2D induced metric on the space-like foliation of $\tilde{\cal H}$. In \rf{gDNSb} $\mathcal{L}_h$ defines the Lie derivative along the vector $h^a$; 
${}^2 D_a$ is the covariant, w.r.t. the 2D induced metric $q_{ab}$, derivative; $T_{ab}$ denotes the EMT of matter fields. For the rest of quantities and symbols entering eq. \rf{gDNSb} we use the notation of \cite{Gourgoulhon:2005ch}. For instance, the ``surface gravity'' 1-form is determined in \cite{Gourgoulhon:2005ch} as
\be
\kappa^{(\mathbf{l})}_a = -\left(\d^b_a - q^b_a\right)k_c\nabla_b l^c .
\la{kappa1def}
\ee
The definitions of the remaining quantities and operations will be given as needed when comparing equation \rf{gDNSb} with the obtained in the main text equation \rf{GCalim}.

To restate eq. \rf{gDNSb} as eq. \rf{GCalim}, we will introduce the orthonormal basis, related to the vectors $(\mathbf{h},\mathbf{m})$:
\be
h^a = \l \hat{u}^a, \qquad m^a=\l\hat{n}^a, \qquad  \l^2 = -2C.
\la{hm2un}
\ee
Apparently,
\be
\hat{u}^a\hat{u}_a = -1, \qquad \hat{n}^a \hat{n}_a = 1, \qquad \hat{u}^a \hat{n}_a = 0 .
\la{tilunorhon}
\ee
Then, in terms of $\hat{u}^a$ and $\hat{n}^a$, the null-vectors $(\mathbf{l},\mathbf{k})$ become
\be
l^a = \frac{\l}{2}\left(\hat{u}^a + \hat{n}^a \right),\qquad k^a = \frac{1}{\l}\left(\hat{u}^a-\hat{n}^a \right),
\la{lk2un}
\ee
so that
\be
\hat{u}^a = \frac{1}{\l}l^a+ \frac{\l}{2}k^a, \qquad \hat{n}^a = \frac{1}{\l}l^a - \frac{\l}{2}k^a .
\la{hatunlk}
\ee

Now, we have to write down eq. \rf{gDNSb} in the basis of $(\mathbf{u},\mathbf{n})$. For $\W_a^{(\mathbf{l})}$ (see eq. (3.21) in Ref. \cite{Gourgoulhon:2005ch}) we get
\be
\Omega_a^{(\mathbf{l})} \equiv \frac{1}{k^e l_e}\, q^b_a\, k_c\nabla_b l^c=q^b_a\, \l^{-1}\pa_b\l - \Omega_a^{(\hat{\mathbf{n}})},
\la{Oml}
\ee
where $\Omega_a^{(\hat{\mathbf{n}})}$ has the same structure as $\Omega_a^{(\mathbf{l})}$ with $\mathbf{l} \ra \hat{\mathbf{n}}$ replacement. Getting eq. \rf{Oml}, we have used the identities $\hat{u}_a\nabla_b \hat{u}^a = \hat{n}_a\nabla_b \hat{n}^a = 0$ and $\hat{u}_a\nabla_b \hat{n}^a = -\hat{n}_a\nabla_b  \hat{u}^a$. The first on the r.h.s. of \rf{Oml} term is equal, on the hypersurface $\tilde{\cal H}$, to $\hat{v}_a \equiv q^c_a\hat{u}^b\nabla_b \hat{u}_c$, so that $\Omega_a^{(\mathbf{l})}= \hat{v}_a-\Omega_a^{(\hat{\mathbf{n}})}$. (It  directly follows from eq. \rf{hatunlk}, and eqs. (4.22), (4.24), (4.25), (4.28) and (4.29) of Ref. \cite{Gourgoulhon:2005ch}.) Therefore, with such an identification, the first term on the l.h.s. of \rf{gDNSb} becomes
 \be
q^c_a\mathcal{L}_h\Omega^{(\mathbf{l})}_c=q^c_a\mathcal{L}_h(\hat{v}_c - \Omega_c^{(\hat{\mathbf{n}})}) = q^c_a\mathcal{L}_h\hat{v}_c - q^c_a\mathcal{L}_h \Omega_c^{(\hat{\mathbf{n}})}=\l\left[q^c_a\mathcal{L}_{\hat{\mathbf{u}}}\hat{v}_c- q_a^c\mathcal{L}_{\hat{\mathbf{u}}}\Omega_c^{(\hat{\mathbf{n}})}\right].
\ee

The second term on the l.h.s. of \rf{gDNSb} is
\be
\theta^{(\mathbf{h})}\Omega_a^{(\mathbf{l})} \equiv q^{ab}\nabla_ah_b \left(\hat{v}_a - \Omega_a^{(\hat{\mathbf{n}})}\right)=\Big| h_b=\l \hat{u}_b \Big|=\l\, \theta^{(\hat{\mathbf{u}})}  \left(\hat{v}_a- \Omega_a^{(\hat{\mathbf{n}})} \right).
\ee
Here we have used $q^{ab} \hat{u}_b=0$ and the following definition for $\theta^{(\mathbf{h})}=q^{ab}\nabla_a h_b$ and $\theta^{(\hat{\mathbf{u}})}\equiv \theta^{(\mathbf{h}\ra \hat{\mathbf{u}})}$.

Let's turn to the r.h.s. of eq. \rf{gDNSb}. In the first term on the r.h.s. we meet the scalar product of vectors $\kappa^{(\mathbf{l})}_a$ and $h^a$:
\be
\langle\kappa^{(\mathbf{l})},\mathbf{h}\rangle \equiv h^a\kappa_a=  -\l \hat{u}^b k_a\nabla_b l^a=\hat{u}^a\nabla_a \l - \l\hat{u}^b\hat{u}^a\nabla_b\hat{n}_a,
\ee
where we have used the definition of the ``surface gravity'' one-form \rf{kappa1def}, eqs. \rf{hm2un}, \rf{lk2un}, and the identity $\hat{u}_a\nabla_b \hat{n}^a = -\hat{n}_a\nabla_b  \hat{u}^a$. Hence, the action of the 2D covariant derivative  on $\langle\kappa^{(\mathbf{l})},\mathbf{h}\rangle$ turns into
\be
{^2}D_a\langle\kappa^{(\mathbf{l})},\mathbf{h}\rangle ={^2}D_a\left(\hat{u}^b\pa_b\l \right) -\left(\hat{u}^b\hat{u}^c\nabla_b\hat{n}_c \right)\,\left[{^2}D_a\l\right]-\l\, \left[{^2}D_a\left( \hat{u}^b\hat{u}^c\nabla_b\hat{n}_c \right)\right].
\ee

The second term on the r.h.s. of \rf{gDNSb} contains
\be
\s_{ba}^{(\mathbf{m})} = \Th_{ba}^{(\mathbf{m})}-\frac{q_{ba}}{2}\th^{(\mathbf{m})} = \Th_{ba}^{(\l\hat{\mathbf{n}})}-\frac{q_{ba}}{2}\th^{(\l\hat{\mathbf{n}})} = \l\s_{ba}^{(\hat{\mathbf{n}})}.
\la{sm2sn}
\ee
Here we have used the definitions $\Th_{ab}^{(\mathbf{m})}  = q^{c}_a q^{d}_b\nabla_c m_d$, $\theta^{(\mathbf{m})}=\Tr \,\Th_{ab}^{(\mathbf{m})}$, and the corresponding relation from \rf{hm2un}. Then, the action of ${^2}D_a$ on $\s_{ab}^{(\mathbf{m})} $ results in
\be
{^2}D^b\s_{ba}^{(\mathbf{m})}=\l\left(\s_{ba}^{(\hat{\mathbf{n}})}\hat{v}^b + {^2}D^b\s_{ba}^{(\hat{\mathbf{n}})}\right).
\ee

Next, for $\theta^{(\mathbf{m})}$ and ${^2}D_a \theta^{(\mathbf{m})}$ we have
\be
\th^{(\mathbf{m})} = \l\th^{(\hat{\mathbf{n}})},\quad {^2}D_a\th^{(\mathbf{m})} =\l\left(\th^{(\hat{\mathbf{n}})}\hat{v}_a + {^2}D_a\th^{(\hat{\mathbf{n}})}\right).
\ee
For the fourth term on the r.h.s. of \rf{gDNSb}, on account of $C=-\l^2/2$ and the definition of $\theta^{(\mathbf{k})}$, we get
\be
\theta^{(\mathbf{k})}\, {}^2D_aC =-\frac12q^{ab}\nabla_b k_a \,{}^2D_a\l^2 =q^c_a\pa_c\l\left(\th^{(\hat{\mathbf{n}})}-\th^{(\hat{\mathbf{u}})}\right) =\l \left(\th^{(\hat{\mathbf{n}})}-\th^{(\hat{\mathbf{u}})}\right)\hat{v}_a.
\ee

Finally, taking into account \rf{hm2un} in the last term of \rf{gDNSb}, and dividing  both sides on $\l$, we arrive at
\be
\begin{split}
	q_a^c\mathcal{L}_{\mathbf{\hat{u}}}(-\Omega_c^{(\hat{\mathbf{n}})}) &+ \th^{(\hat{\mathbf{u}})}(- \Omega_a^{(\hat{\mathbf{n}})}) =  q_a^b\pa_b\left(\frac{\th^{(\hat{\mathbf{n}})}}{2}-\hat{u}^c\hat{u}^d\nabla_c\hat{n}_d\right) - {^2}D^b\s_{ba}^{(\hat{\mathbf{n}})} + 8\pi q_a^cT_{cd}\hat{n}^d\\
	&-\left(q_{ba}\hat{u}^c\hat{u}^d\nabla_c\hat{n}_d+\frac{\th^{(\hat{\mathbf{n}})}}{2}q_{ba}+\s_{ba}^{(\hat{\mathbf{n}})}\right)\hat{v}^b - q_a^c\mathcal{L}_{\mathbf{\hat{u}}}\hat{v}_c + \l^{-1}q_a^c\pa_c(\hat{u}^b\pa_b\l)\,.
\end{split}
\la{gDNSb1}
\ee

It is straightforward to verify that $\l^{-1}q_a^c\pa_c(\hat{u}^b\pa_b\l) = q^c_a\mathcal{L}_{\hat{\mathbf{u}}}\hat{v}_c$, so that two last terms on the r.h.s. of \rf{gDNSb1} cancel each other. Therefore, under the following replacements in eq. \rf{gDNSb1},
\be
-\Omega_c^{(\hat{\mathbf{n}})} \leadsto \hat{\Omega}_a, \quad \theta^{(\hat{\mathbf{n}})} \leadsto \hat{\theta},\quad -\hat{u}^b\hat{u}^a\nabla_b\hat{n}_a \leadsto \hat{g}, \quad \s_{ba}^{(\hat{\mathbf{n}})} \leadsto \hat{\s}_{ba}, \quad q_{ab} \leadsto \g_{ab},
\ee
with omitting hats over the basic vectors $(\hat{\mathbf{u}},\hat{\mathbf{n}})$, we get the following equation outside of the horizon:
\be
\g_a^b\pa_b\left(\frac{\hat{\th}}{2}+\hat{g}\right) - {^2}D^b\hat{\s}_{ba} +\left(\g_{ac}\left(\hat{g}-\frac{\hat{\th}}{2}\right)-\hat{\s}_{ac}\right)\hat{v}^c+ 8\pi \g_a^cT_{cd}{n}^d=\g_a^c\mathcal{L}_{\mathbf{{u}}}\hat{\Omega}_c + \hat{\Omega}_a\,\hat{\theta}  .
\ee
Comparing this equation to eq. \rf{GCa}, we note that, modulo terms $\g^c_a\hat{\Omega}^b(\nabla_bu_c - \nabla_cu_b)$, two equations coincide. However, in the particular representation for $u_a$ and $n_a$ used in \cite{Gourgoulhon:2005ch,Gourgoulhon:2005ng}, 
\be
u_a=-Uh^b_a\nabla_b\tau, \qquad n_a = N\nabla_a\r, 
\la{nuGJ}
\ee
with scalar functions $U$, $\t$, $N$ and $\r$, it is easy to verify $\g^c_a\hat{\Omega}^b(\nabla_bu_c - \nabla_cu_b)=0$. 

To sum up, we have proved the equivalence of the generalized DNS equation of Ref. \cite{Gourgoulhon:2005ch} (eq. \rf{gDNSb}) to the generalization of the DNS equation (eq. \rf{GCa}) obtained within the Membrane Approach of \cite{Parikh:1997ma}.

\bsk
\section{The Null-Hypersurface Limit of the Gourgoulhon's generalization of the DNS equation}

It was claimed in Ref. \cite{Gourgoulhon:2005ch} that in the null-hypersurface limit eq. \rf{gDNSb} turns into the standard version of the DNS equation. In this Appendix we will take this limit for the equivalent to \rf{gDNSb} equation, 
\be
\g_a^c\mathcal{L}_{\mathbf{{u}}}\hat{\Omega}_c + \theta^{({\mathbf{u}})} \hat{\Omega}_a =  \g_a^b\pa_b\left(\frac{\th^{({\mathbf{n}})}}{2}+\hat{g}\right) - {^2}\mathbf{D}^b\s_{ba}^{({\mathbf{n}})} + 8\pi \g_a^cT_{cd}{n}^d
	+\left(\g_{ba}\left(\hat{g}-\frac{\th^{({\mathbf{n}})}}{2}\right)-\s_{ba}^{({\mathbf{n}})}\right){v}^b \,.
\la{gDNSb2}
\ee

The null-hypersurface limit is realized by eqs. \rf{rel1} and \rf{leq}, that are
\[
\lim_{\a\rightarrow0}\a u^a = l^a, \quad \lim_{\a\rightarrow0}\a n^a = l^a,\quad l^b\nabla_b l^a = g_\mathcal{H}\,l^a.
\]
In this limit, geometric quantities become divergent (if so) as inverse degrees of $\a$. We will regularize them by turning to finite on the horizon variables, which are
\be
\theta^{(\mathbf{u})} = \a^{-1}\bar{\theta},\quad \theta^{(\mathbf{n})} = \a^{-1}\bar{\theta}, \quad \hat{g} =\a^{-1}\bar{g},\quad \s^{(\mathbf{n})}_{ab} = \a^{-1}\bar{\s}_{ab},\quad \Omega_a = \bar{\Omega}_a.
\la{regthungs}
\ee
The bar over quantities means their regularity (finiteness) in the $\a \ra 0$ limit.

Near the event horizon (on the stretched horizon, where $\a$ is small but not equal to zero) the vectors $u^a$ and $n^a$ admit the form
\be
u^a = \a^{-1}l^a + \a \d^a, \;\; n^a = \a^{-1}l^a + \a \b^a .
\la{undb}
\ee
By use of the orthonormality relations between $u^a$ and $n^a$, similar to eqs. \rf{tilunorhon}, up to the 2nd order in $\a$, we get
\be
l^a \d_a=-\fr12+\cO(\a^2),\quad l^a \b_a=\fr12+\cO(\a^2),
\la{ldbrel}
\ee
from which it follows $l^a \d_a=-l^a \b_a$. Also, due to the orthogonality of $u^a$ and $n^a$ to the 2D induced metric $\g_{ab}$, the same property is translated onto the vectors $\d^a$ and $\b^a$. The representation \rf{undb} allows one to recover the regularized on the horizon $\theta^{(\mathbf{u})}$, $\theta^{(\mathbf{n})}$, $\s^{(\mathbf{n})}_{ab}$, and $\hat{g}$. For instance,
\be
\theta^{(\mathbf{u})} \equiv \g^{ab}\nabla_bu_a=\a^{-1}\theta^{(\mathbf{l})} + \mathcal{O}(\a),
\ee
\be
\theta^{(\mathbf{n})} \equiv \g^{ab}\nabla_bn_a =\a^{-1}\theta^{(\mathbf{l})} + \mathcal{O}(\a),
\ee
where we have used the orthogonality of $\d^a$ and $\b^a$ to $\g_{ab}$, and the definition of $\theta^{(\mathbf{l})}\equiv \g^{ab}\nabla_b l_a$. Comparing the obtained results to \rf{regthungs}, we conclude that
\be
\bar{\theta} = \theta^{(\mathbf{l})} + \mathcal{O}(\a^2).
\ee
Next, for $\s^{(\mathbf{n})}_{ab}$, we arrive at
\be
\s^{(\mathbf{n})}_{ab} = \a^{-1}\s^{(\mathbf{l})}_{ab}  + \mathcal{O}(\a) \quad \leadsto \quad \bar{\s}_{ab} = \s^{(\mathbf{l})}_{ab} +\mathcal{O}(\a^2).
\ee

Computations of $\hat{g}$ are more involved. Here we have
\be
\begin{split}
	&\hat{g} = -u^au^b\nabla_bn_a = -(\a^{-1}l^a + \a\d^a)(\a^{-1}l^b + \a\d^b)\nabla_bn_a\\
	&=-(\a^{-2}l^al^b\nabla_bn_a + l^a\d^b\nabla_bn_a + l^b\d^a\nabla_bn_a + \a^2\d^a\d^b\nabla_bn_a),
\end{split}
\ee
after that we have to use eqs. \rf{ldbrel}, together with the relations $l^b\nabla_bl^a=g_\mathcal{H}\,l^a$, $l^al_a=0$ and $l^a\nabla\b_a =-\b^a\nabla l_a$. As a result, we arrive at
\be
\hat{g}=\a^{-1}\left(g_\mathcal{H}-l^b\pa_b\ln\a + \mathcal{O}(\a^2)\right),
\ee
that gives
\be
\bar{g}=g_\mathcal{H}-l^b\pa_b\ln\a + \mathcal{O}(\a),
\ee
with the surface gravity $g_{\cal H}$. Also, one can see that ${v}_a=\g_a^cu^b\nabla_bu_c=v_a^{\text{ finite}}+\cO(\a^2)$.

Now, in terms of the regular on the horizon variables (up to the leading order in $\a$), eq. \rf{gDNSb2} turns out to be
\be
\begin{split}
&\g_a^c\mathcal{L}_l\Omega_c + \th^{(\mathbf{l})}\Omega_c \cong\g^b_a\pa_b\left(g_\mathcal{H}-l^b\pa_b\ln\a+\frac{\th^{(\mathbf{l})}}{2}\right) -\,{^2}\mathbf{D}^b\s^{(\mathbf{l})}_{ba} + 8\pi\g^c_aT_{cd}l^d\\
	&+\left(\s^{(\mathbf{l})}_{ba}-\left(\frac{\th^{(\mathbf{l})}}{2}+g_\mathcal{H}-l^b\pa_b\ln\a\right)\g_{ba}\right)\g^{bc}\pa_c\ln\a +\left(\left(g_\mathcal{H}-l^b\pa_b\ln\a-\frac{\th^{(\mathbf{l})}}{2}\right)\g_{ba}-\s^{(\mathbf{l})}_{ba}\right)v^b\,.
\la{gDNSb3}
\end{split}
\ee
Comparing the so obtained eq. \rf{gDNSb3} with the original DNS equation
\be
\begin{split}
	\g_a^c\mathcal{L}_l\Omega_c + \th^{(\mathbf{l})}\Omega_c &=\g^b_a\pa_b\left(g_\mathcal{H}+\frac{\th^{(\mathbf{l})}}{2}\right) -\,{^2}\mathbf{D}^b\s^{(\mathbf{l})}_{ba} + 8\pi\g^c_aT_{cd}l^d,
\la{DNSclas}
\end{split}
\ee
we derive the same conditions on the regularization function $\a$ as in the main text of the paper (cf. eqs. \rf{rel2}):
\be
 \left(\left(g_\mathcal{H}-\frac{\th^{(\mathbf{l})}}{2}\right)\g_{ab}-\s^{(\mathbf{l})}_{ab}\right)v^b\overset{\mathcal{H}}{=}\left(\left(\frac{\th^{(\mathbf{l})}}{2}+g_\mathcal{H}\right)\g_{ab}-\s^{(\mathbf{l})}_{ab}\right)\g^{bc}\pa_c\ln\a, \;\;\;\; l^b\pa_b\ln\a \overset{\mathcal{H}}{=} 0\,.
 \la{condition}
\ee
Here the symbol ``$\overset{\mathcal{H}}{=}$'' is reserved for computing quantities and their derivatives on the horizon.

Now, let's fix the particular choice of $u^a$ and $n^a$ vectors from Appendix B (eq. \rf{hatunlk}):
\be
\hat{u}^a = \frac{1}{\l}l^a+ \frac{\l}{2}k^a, \qquad \hat{n}^a = \frac{1}{\l}l^a - \frac{\l}{2}k^a ,
\la{hatunlk1}
\ee
and let's assume that $l^a$ is the event horizon null-generator, i.e., $l^b\nabla_b l^a = g_\mathcal{H}\,l^a$. Since the parameter $\l$ in \rf{hatunlk1} is free, we can identify it with the regularization function $\a$. Then, since as an outcome of Appendix B we have obtained $\hat{v}_a=\g_a^c \l^{-1}\pa_b \l$, we will have $v^b\overset{\mathcal{H}}{=}\g^{bc} \pa_c \ln \a$ on the horizon. 

The second condition of \rf{condition} is always satisfied with the given choice of the regularization function. The rest of \rf{condition} is
\be
\theta^{(\mathbf{l})}\g^c_b\pa_c\ln{\a}\overset{\mathcal{H}}{=} 0,
\ee
which requires either space-time configurations with the trivial vector $v_a$, or, in the case of finite-valued $v_a$ on the horizon, the trivial on the horizon expansion $\theta^{(\mathbf{l})}\overset{\mathcal{H}}{=} 0$. For example,  
\begin{itemize}
	
	\item for the Kerr BH,  $l^b\pa_b\ln\hat{\a} = 0, \; \g^c_b\pa_c\ln\hat{\a}\overset{\mathcal{H}}{=}\text{finite}\neq0$, and $\th^{(\mathbf{l})}\overset{\mathcal{H}}{=}0$,  
	\item for the Schwarzschild BH, $l^b\pa_b\ln\hat{\a} = 0, \; \g^c_b\pa_c\ln\hat{\a}=0$, and $\th^{(\mathbf{l})}\overset{\mathcal{H}}{=}0$.
\end{itemize}

Now, it becomes clear that the claim of \cite{Gourgoulhon:2005ch} on the equivalence of eq. \rf{gDNSb} to eq. \rf{DNSclas} in the horizon (null-hypersurface) limit is based on the possibility, within the approach of \cite{Gourgoulhon:2005ch,Gourgoulhon:2005ng}, to fix $C=0$ and $h^a=m^a=l^a$ (see eq. \rf{gDNSb}). Put it differently, in terms of the vector $v_a$ and the regularization function $\a$, in Ref. \cite{Gourgoulhon:2005ch} it is silently supposed that we can always fix $v_a=0$ in the $\a \ra 0$ limit. However, as we have convinced on the example of the Kerr Black Hole, it is not always possible for a general Black Hole spacetime metric.

\bsk\bsk


\end{document}